\def\n{\nu}

\def\l{\lambda}

\documentstyle{elsart}
\begin{document}
\begin{frontmatter}
\title{Implications of the Measurements of $\bf U_{e3}$ to Theory
\thanksref{talk}}
\thanks[talk]{Talk at the  International Workshop  NuFACT'01,
 Tsukuba, Japan (May 2001)}
\author{Morimitsu TANIMOTO}
\address{Department of Physics, Niigata University, Ikarashi 2-8050,
  Niigata, JAPAN}
%%%%%%%%%%%%%%%%%%%%%%%%%%%%%%%%%%%%%%%%%
\begin{abstract}
The predictions of $U_{e3}$ are  discussed. Typical models which
lead to the large, the seizable and the tiny  $U_{e3}$ are also studied.
\end{abstract}
\end{frontmatter}
%%%%%%%%%%%%%%%%%%%%%%%%%%%%%%%%%%%%%%%%%
%%%%%%%%%%%% Section 1  %%%%%%%%%%%%%%%%%
%%%%%%%%%%%%%%%%%%%%%%%%%%%%%%%%%%%%%%%%%
\section{Introduction}
  The measurement of the neutrino mixing $U_{e3}$ is the main target
in the long baseline experiments of the neutrino oscillations.
The magnitude of the $U_{e3}$ will give a big impact on the model building.
 Furtheremore, the CP violation measure  $J_{CP}$ and the lepton flavor
 violation $Br(\mu \rightarrow e\gamma)$  depend on  $U_{e3}$.

 We assume that oscillations need only account for 
 the solar neutrinos  and atmospheric neutrinos  data in the three families.
Then, the atomospheric  neutrino data constrain
$U_{\mu 3}\simeq U_{\tau 3}\simeq  1/\sqrt{2}$, on the other
hand, there are a few options of the mixing:
$U_{e2}\simeq 1/\sqrt{2} ,\  \l^{2\sim 3} \ (\l\simeq 0.2)$,
which depend on the solar neutrino solutions.
Therefore, there are two possible mixing matrices:
the bi-maximal and the single maximal ones.
%%%%%%%%%%%%%%%%%%%%%%%%%%%%%%%%%%%%%%%%%%%%%%%%%%%%%
%%%%%%%%%%%%%%%%%%%%%%%%%%%%%%%%%%%%%%%%
%%%%%%%%%%%%%%%%%%%%%%%%%%%%%%%%%%%%%%%%%

How large is $U_{e3}$?
The Choose experiments constrain $U_{e3}< 0.16$ \cite{Choose}.
 Atmospheric neutrino and  solar neutrino  data 
also constraint  $U_{e3}$ \cite{bound}.
Let us consider the case of the quark mixings.
 CKM mixings are given in terms of  $\l$
  in the Wolfenstein parametrization as follows:
$|V_{ub}|\simeq \l^{3} , \ |V_{cb}|\simeq \l^2 , \ |V_{us}|\simeq \l$,
which are  related with quark mass ratios.
Since the MNS \cite{MNS}  mixings are much different from CKM mixings,
 it is very difficult to a reliable  prediction  of $U_{e3}$.
%%%%%%%%%%%%%%%%%%%%%%%%%%%%%%%%%%%%%%%%%%%%%%
%%%%%%%%%%  Section 2   Model     %%%%%%%%%%%%%%%%%%%%%
%%%%%%%%%%%%%%%%%%%%%%%%%%%%%%%%%%%%%%%%%%%%%%
%%%%%%%%%%%%%%%%%%%%%%%%%%%%%%%%%%%%%%%%%%%%%%%%%%%%%%%%
\section{How to predict $U_{e3}$}
Since there is a preferred basis given by the underlying theory of the model,
the MNS mixing matrix is given as follows:
\begin{eqnarray}
U_{MNS}=L_E^\dagger L_\n \ , \qquad
L_E^\dagger M_E  R_E = M_E^{\rm diag} \ , \qquad
             L_\nu^T M_\nu L_\nu = M_\nu^{\rm diag} .
 \end{eqnarray}
%%%%%%%%%%%%%%%%%%%%%%%%%%%%%
\noindent Taking  $(L_E)_{ij}=L_{ij}$ and  $(L_\nu)_{ij}=N_{ij}$, 
we get $U_{e2}$,  $U_{\mu 3}$ and  $U_{e3}$ as
\begin{eqnarray}
U_{e2}&=&L_{11}^* N_{12}+L_{21}^* N_{22}+L_{31}^* N_{32} \ , \quad
 U_{\mu 3}=L_{12}^* N_{13}+L_{22}^* N_{23}+L_{32}^* N_{33} \ , \nonumber \\
 U_{e3}&=&L_{11}^* N_{13}+L_{21}^* N_{23}+L_{31}^* N_{33} \ .
\end{eqnarray} 
We show a few examples with $U_{\mu 3}\simeq U_{\mu 3}\simeq 1/\sqrt{2}$.
The first case is 
$(1):    N_{23}\simeq  N_{33}\simeq 1/\sqrt{2},\
   L_{ij}\ll 1 (i\not =j)$.
Then, we get 
$U_{e 2}=N_{12}+(L_{21}^*-L_{31}^*)/\sqrt{2}$ and  
$ U_{e 3}=N_{13}+(L_{21}^*+L_{31}^*)/\sqrt{2}$.
If $ L_{21}\gg L_{31}, N_{12}, N_{13}$, we get $U_{e 2}\simeq  U_{e 3}$, 
on the other hand, 
if $N_{12}\gg  L_{21}, L_{31}, N_{13}$, we get $U_{e 2}\gg U_{e 3}$.

%%%%%%%%%%%%%%%%%%%%%%%%%%%%%%%%%%%%%%%%%%%%%
The second case is  
$(2):L_{22}\simeq  L_{32}\simeq 1/\sqrt{2} ,\
   N_{ij}\ll 1 (i\not =j)$.
Then, $U_{e 2}=N_{12}+L_{21}^*$ and 
$ U_{e 3}=N_{13}+L_{21}^* N_{23}+L_{31}^*$ are obtained.
If $N_{12}\simeq  N_{23}\simeq L_{21}\simeq \l$, we get 
  $U_{e 2}\simeq 2\l$ (LMA-MSW) and   $ U_{e 3}\simeq \l^2$.
The last case is 
$(3):$  NNI-form texture \cite{NNI} of $M_E$. 
In this basis, we can give
\begin{eqnarray}
 {\rm M_U}= m \left ( \matrix{0 & a_u& 0 \cr 
  a_u&  0 & b_u \cr 0 & b_u & c_u\cr }
             \right ) \ , \qquad\quad
{\rm M_D}= m \left ( \matrix{0 & a_d & 0\cr 
 c_d & 0 & b_d \cr 0 & d_d & e_d\cr } \right ) \ .
\end{eqnarray} 
Using the SU(5) relation  $M_E=M_D^T$,
%%%%%%%%%%%%%%%%%%%%%%%%%%%%%%%%%%%%%%%%%%%%%%%%%%%%%%%%%%%%
we obtain
\begin{eqnarray}
 L_E^{\dagger}=  \left( \matrix{
1 & -\sqrt{\frac{m_e^{}}{m_\mu^{}}}\frac{y^2}{yz}
  &  \sqrt{\frac{m_e^{}}{m_\mu^{}}}\frac{\sqrt{1-y^4}}{yz} \cr
-\frac{1}{yz}\sqrt{\frac{m_e^{}}{m_\mu^{}}}\frac{m_\mu^{}}{m_\tau^{}}
  &  y^2 & -\sqrt{1-y^4} \cr
\frac{y}{z\sqrt{1-y^4}}\frac{m_\mu^{}\sqrt{m_e^{}m_\mu^{}}}{m_\tau^2}
  &  \sqrt{1-y^4} & y^2  }  \right) \ ,
\end{eqnarray}
\noindent
where $y$, $z$  are free parameters.
 Assuming $U_{MNS}\simeq L_E^\dagger$, which corresponds to  
 $ N_{ij}\ll 1 (i\not =j)$, we predict
%%%%%%%%%%%%%%%%%%%%%%%%%%%%%%%%%%%%%%%%%%%%%%%%%%%%%%%%%%%%
\begin{eqnarray} 
\cot \theta_{\mu\tau}\equiv
\left|\frac{U_{\tau 3}}{U_{\mu 3}}\right|
=
\left|\frac{U_{\mu 2}}{U_{\mu 3}}\right|
=
\left|\frac{U_{e 2}}{U_{e 3}}\right|  , \quad
 \cot \theta_{\mu\tau}= \frac{m_b}{m_s} \ \left| \ 
 V_{cb} + \sqrt{\frac{m_c}{m_t}} e^{-i\theta} \  \right|  ,
\end{eqnarray}
\noindent which is testable in the future.
%%%%%%%%%%%%%%%%%%%%%%%%%%%%%%%%%%%%%%%%%%%%%%%%%%%%%%%%
%%%%%%%%%%%%%%%%%  Section 3 %%%%%%%%%%%%%%%%%%%%%%%%%%%
%%%%%%%%%%%%%%%%%%%%%%%%%%%%%%%%%%%%%%%%%%%%%%%%%%%%%%%%
\section{Large  $ U_{e3}$ or Tiny  $U_{e3}$}
Let us consider models which predict the large  $U_{e3}\simeq  O(\l \sim 1)$.
The typical model is the anarchy \cite{Anarchy}. 
In this model, 
there is no suppression in the Yukawa couplings, and so all elements
are the same order.
  Then,  one predicts  $U_{e3}\sim O(1)$, which should be observed very soon.

Next, consider models with   seizable  $U_{e3}\simeq  O(\l^3 \sim \l)$.
The typical one is  models with U(1) flavor symmetry \cite{U1}
 or the conformal fixied point (CFT) theory \cite{CFP}.
In these models, the Yukawa couplings are given as 
  $Y_{ij}=\epsilon_{L i}\epsilon_{R i}$ with 
$\epsilon_{L(R)i}=(M_C/M_X)^{a_{L(R)i}}$.
Since $\theta_{Lij}\simeq \epsilon_{Li}/\epsilon_{Lj}$
 and $m_i/m_j=(\epsilon_{Li} \epsilon_{Ri})/(\epsilon_{Lj} \epsilon_{Rj})$,
we get 
{$U_{e3}=U_{e2}\times U_{\tau 2}\simeq (1/\sqrt{2})U_{e2}$.
Models with the  non-abelian flavor symmetry also predict
  $\rm U_{e3}=O(\l\sim \l^2)$ \cite{Democratic}.

Let us consider  the case of the tiny $U_{e3}\ll  O(\l^3)$.
Typical one is the generation of radiative neutrino masses \cite{Radi}
as follows:
\begin{eqnarray}
 M_\nu\sim 
\left (\matrix{ 0 &  1 &  1 \cr 1 & 0 & \epsilon \cr 1 & \epsilon & 0 \cr }
 \right )
 \ , \qquad\qquad
 U_\nu= \left(\matrix{\frac{1}{\sqrt{2}}&\frac{1}{\sqrt{2}}&0\cr    
- \frac{1}{2} &  \frac{1}{2}& \frac{1}{\sqrt{2}} \cr
 \frac{1}{2} & -\frac{1}{2} &  \frac{1}{\sqrt{2}} \cr } \right )  \ .
\end{eqnarray}
\noindent 
In this model, one gets the inverse  mass hierarchy
 $|m_1|\simeq |m_2| \gg |m_3|$ and  $\rm U_{e3}\leq 10^{-4}$.  
%%%%%%%%%%%%%%%%%%%%%%%%%%
%%%%%%%%%%%%%%%%  Summary  %%%%%%%%%%%%% 
%%%%%%%%%%%%%%%%%%%%%%%%%%%%%%%%%%%%%%%%%%%%%%%
%%%%%%%%%%%%%%%%%%%%%%%%%%%%%%
\section{Summary}
There are many models which predict  $U_{e3}$.
The large $U_{e3}$ and the tiny one can be  predicted  as well as 
the seizable one.
In order to test the models precisely, we need testable sum rules 
among  $V_{CKM}$ and $U_{MNS}$  elements \cite{NNI,Relation}.
%%%%%%%%%%%%%%%%%%%%%%%%%%%%%%%%%%%%%%%%
For example, there are beautiful relations  \cite{Relation} as follows: 
\begin{eqnarray} 
\cot \theta_{\mu\tau}\equiv
\left|\frac{U_{\tau 3}}{U_{\mu 3}}\right|=
\left|\frac{U_{\mu 2}}{U_{\mu 3}}\right|=
\left|\frac{U_{e 2}}{U_{e 3}}\right| \ , \quad\quad
 \cot \theta_{\mu\tau}= \frac{m_b}{m_s} \ \left| \ V_{cb}  \  \right| \ ,
\end{eqnarray}
%%%%%%%%%%%%%%%%%%%%%%%%%%%%%%%%%%%%%%%%%
which can be tested by LBL experiments in the future.
The measurement of $U_{e3}$ will give a big impact on the model building.


\begin{thebibliography}{9}
\bibitem{Choose}
  The CHOOZ Collaboration, M.~Apollonio et al., Phys.~Lett.  {\bf B420}
 (1998) 397.
\bibitem{bound}
 M. C. Gonzalez-Garcia, M. Maltoni, C. Pena-Garay and J.W.F. Valle,
 Phys. Rev. {\bf D63} (2001) 033005.
\bibitem{MNS}
  Z.~Maki, M.~Nakagawa and S.~Sakata, Prog.~Theor.~Phys. {\bf 28} (1962) 870.
\bibitem{NNI}
 L. Lavoura, G. C. Branco and F. Mota,  Phys. Rev. {\bf D39} (1989) 3443; \\
 E. Takasugi, Prog. Theor. {\bf 98} (1997) 177; \\
 E. Takasugi and M. Yoshimura, Prog. Theor. {\bf 98} (1997) 1313;\\
  N. Okamura and M. Tanimoto, Prog. Theor. {\bf 105}  (2001) 459.
\bibitem{Anarchy}
 L.  Hall,  H. Murayama,  N. Weiner, Phys.~Rev.~Lett. {\bf 84} (2000) 2572; \\
  N. Haba and H. Murayama,  Phys.~Rev. {\bf D63} (2001) 053010.
\bibitem{U1}
  J. K.~Elwood, N.~Irges and P.~Ramond, Phys.~Rev.~Lett. {\bf 81}
  (1998) 5064.
\bibitem{CFP}
A.E.  Nelson and  M. J. Strassler,  JHEP{\bf 0009} (2000) 030.
\bibitem{Democratic}
  M.~Fukugita, M.~Tanimoto and T.~Yanagida, Phys.~Rev. {\bf D57} 
  (1998) 4429;\\
  M.~Tanimoto, Phys.~Rev. {\bf D59} (1999) 017304;\\
  M.~Tanimoto, T.~Watari and T.~Yanagida, Phys.~Lett. {\bf B461}
  (1999) 345.
\bibitem{Radi} 
  C.~Jarlskog, M.~Matsuda, S.~Skadhauge and M.~Tanimoto, Phys.~Lett. 
  {\bf B449} (1999) 240;\\
  P. H.~Frampton and S.~Glashow, Phys.~Lett. {\bf B461} (1999) 95.
\bibitem{Relation} 
 C.H.~Albright and S.M.~Barr, Phys.~Lett. {\bf B452} (1999) 287;\\
 M.~Bando, T.~Kugo and K. Yoshioka, Phys.~Lett. {\bf B483} (2000) 163;\\
 Prog.~Theor.~Phys. {\bf 104} (2000) 211;\\
 A. Kageyama, M. Tanimoto and K. Yoshioka, hep-ph/0102006;\\
 see also Okamura and Tanimoto in ref.[3].
\end{thebibliography}
\end{document}